\documentclass[fleqn,usenatbib]{mnras}

\usepackage{newtxtext,newtxmath}

\usepackage[T1]{fontenc}
\usepackage{ae,aecompl}

\usepackage{graphicx}	
\usepackage{amsmath}	
\usepackage{amssymb}	
\usepackage{footnote}
\usepackage{longtable}
\usepackage{amssymb}

\usepackage[authoryear]{natbib}
\bibpunct{(}{)}{;}{a}{}{,}
\label{Bibliography}

\newcommand{\apspr}{APSPR}  

\title[Thermally driven winds in ULXs]
{Thermally driven winds in ULXs}

\author[M. Middleton et al.]
{Matthew J. Middleton$^{1}$ \thanks{E-mail: m.j.middleton@soton.ac.uk},  Nick Higginbottom$^{1}$, Christian Knigge$^{1}$, Norman Khan$^{1}$ \newauthor \& Grzegorz Wiktorowicz$^{2,3,4}$
\\
\\
$^{1}$Department of Physics \& Astronomy, University of Southampton, Southampton, SO17 1BJ, UK\\
$^{2}$National Astronomical Observatories, Chinese Academy of Sciences, Beijing 100101, China\\
$^{3}$School of Astronomy \& Space Science, University of the Chinese Academy of Sciences, Beijing 100012, China\\
$^{4}$ Nicolaus Copernicus Astronomical Center, Polish Academy of Sciences, Bartycka 18, 00-716 Warsaw, Poland
\\
}

\date{Accepted XXX. Received YYY; in original form ZZZ}

\pubyear{2020}

\begin{document}
\label{firstpage}
\pagerange{\pageref{firstpage}--\pageref{lastpage}}
\maketitle

\begin{abstract}
The presence of radiatively driven outflows is well established in ultraluminous X-ray sources (ULXs). These outflows are optically thick and can reprocess a significant fraction of the accretion luminosity. Assuming isotropic emission, escaping radiation from the outflow's photosphere has the potential to irradiate the outer disc. Here, we explore how the atmosphere of the outer disc would respond to such irradiation, and specifically whether unstable heating may lead to significant mass loss via thermally-driven winds. We find that, for a range of physically relevant system parameters, this mass loss may actually switch off the inflow entirely and potentially drive limit-cycle behaviour (likely modulated on the timescale of the outer disc). In ULXs harbouring neutron stars, magnetic fields tend to have a slight destabilizing effect; for the strongest magnetic fields and highest accretion rates, this can push otherwise stable systems into the unstable regime. We explore the prevalence of the instability in a simulated sample of ULXs obtained from a binary population synthesis calculation. We find that almost all neutron star and black hole ULXs with Eddington-scaled accretion rates of $\dot{m}_0 < 100$ should be able to drive powerful outflows from their outer discs. Several known ULXs are expected to lie in this regime; the persistence of accretion in these sources implies the irradiation may be anisotropic which can be reconciled with the inferred reprocessed (optical) emission if some of this originates in the wind photosphere or irradiation of the secondary star.

\end{abstract}






\section{Introduction}
It is accepted that ultraluminous X-ray sources (ULXs) harbour stellar mass compact objects accreting matter from a companion at rates up to many orders of magnitude above the Eddington limit. The strongest evidence in support of this paradigm has been through the detection of pulsations in several well studied systems, indicating that the accretors must be magnetised neutron stars \citep{NSULX_Bachetti_2014, NSULX_Furst_2016, Israel_2017_NSULX_5907,
Tsygankov_2017_NSULX_SMCX3, Doroshenko_2018_NSULX_Swift_J0243,
Carpano_2018_NSULX_NGC300, Sathyaprakash_2019_1313X2, 2020_Rodriguez_ApJ...895...60R}. The same inference can be made for at least one other ULX, in which a cyclotron resonance scattering feature (CRSF) has been discovered (\citealt{2018_Brightman_CRSF,
Middleton_Brightman_2019_M51}). The overall population is expected to be a heterogeneous mix of black holes and neutron stars (e.g. \citealt{Middleton_2017_demographics_from_beaming, King_Lasota_2020_PULX_iceberg_emerges}), with the former predicted to dominate the observed population only at low metallicities \citep{Wiktorowicz_2019_obs_vs_tot}). 

Standard models to explain the accretion flow in ULXs (\citealt{Shakura_1973,Poutanen_2007_ln}) agree that if the Eddington limit is reached locally in the disc (around the spherisation radius), the radiation pressure matches the vertical component of gravity. This leads to a large scale height (H/R$\sim$1) inflow, radial advection and radiatively driven winds. In neutron star ULXs, the nature of the accretion flow depends on the dipole magnetic field and how close the system is to spin equilibrium. Specifically, if the spherisation radius ($r_{\rm sph}$) is much larger than the magnetospheric radius ($r_{\rm M}$), black hole ULXs and neutron star ULXs will look somewhat similar (modulo differences in the inner regions, e.g. \citealt{Mushtukov_2017_opt_thick_env_NS}), although the optical depth through the wind will depend on the dipole magnetic field strength (see \citealt{Middleton_2019_Accretion_plane, 2019_Vasilopoulos_MNRAS.488.5225V}). In such cases, the radiation-driven wind will be launched from the disc region interior to $r_{\rm sph}$ and is expected to remain optically thick out to some photospheric radius, $r_{\rm ph} >r_{\rm sph}$ (\citealt{Poutanen_2007_ln}). 

The accretion power liberated between $r_{\rm ph}$ and $r_{\rm sph}$ will be scattered/reprocessed and emerge from the outer face of the wind at $r_{\rm ph}$. At least some of this radiation must irradiate and heat the outer disc (at $r > r_{\rm ph}$). The actual strength of this irradiation depends on the largely unknown geometry of the radiation field. If the temperature in the irradiated disc atmosphere becomes sufficiently high, the thermal speed of particles can exceed the local escape speed; gas will then become unbound. Such irradiation-driven thermal winds are often invoked (along with magneto-centrifugal contributions) to explain the sub-relativistic winds seen in X-ray binaries (\citealt{Ponti2012_thermalwinds}). Radiation-hydrodynamic (RHD) simulations indicate that the relative mass-loss rate in such outflows can be extreme, carrying away 90\% or more of the available matter (\citealt{Higginbottom17}). Thermal winds may therefore have a profound effect on the long timescale behaviour and appearance of a system (e.g. \citealt{Dubus2019_winds}).

Should irradiation of the outer disc be possible in ULXs (see also \citealt{Sutton2014_irraddisc, Yao_Feng2019_discirrad}), thermal winds may potentially affect the structure of the inner disc by changing the accretion rate feeding it. This, in turn, may change observational signatures, including the peak luminosity (\citealt{King_2009_Beaming}), the spectrum (\citealt{Poutanen_2007_ln}), the appearance of wind-formed features (e.g. \citealt{Pinto_winds_thermalbalance}), the spin-up rate of the neutron star (\citealt{2019Chashkina}), the fast variability properties (\citealt{Middleton_2015ULX_modelpaper}) and the slower variations associated with disc precession (\citealt{Middleton_2018_Lense_Thirring,Middleton_2019_Accretion_plane}). In this paper, we explore the impact of irradiation-driven thermal disc winds on ULXs, and indicate where in parameter space such outflows should be expected.

\section{the model}
We base our model on the description of super-critical accretion provided in \cite{Poutanen_2007_ln}, which describes a classical super-critical, radiation-pressure supported accretion flow with mass-loaded, moderately relativistic outflows. We note that this model does not account for a neutron star surface and dipole magnetic field, both of which will affect the structure of the disc and wind - (see e.g. \citealt{Mushtukov_2017_opt_thick_env_NS}). The flow reaches the local Eddington limit around $r_{\rm sph} \approx \dot{m}r_{\rm in}$ (see also \citealt{Fukue2004}), where lower case $\dot{m}$ indicates the mass accretion rate in units of the Eddington accretion rate, i.e. $\dot{M}/\dot{M}_{\rm Edd}$, and $r_{\rm in}$ is the inner edge of the disc, assumed to be close to the ISCO, (usually taken to be 6 $R_{\rm g}$, although this is not reached for highly magnetised neutron stars) in units of gravitational radii (where $R_{\rm g} = GM/c^{2}$). 

The outer photosphere of the radiation-driven wind is assumed to be located at $r_{\rm ph} \sim (3\epsilon_{\rm w}/\xi \beta)\dot{m}^{3/2}$ (in units of $r_{\rm in}$). Here, $\epsilon_{\rm w}$ is the fraction of the radiative luminosity used in launching the wind (we take $\epsilon_{\rm w}= 0.5$), and $\xi$ and $\beta$ are related to the opening angle and velocity of the wind, respectively (we assume $\xi \simeq \beta \simeq 1$, e.g. \citealt{Poutanen_2007_ln}). 

We assume that the wind photosphere radiates isotropically as a spherical blackbody at the Eddington luminosity (in reality, the photosphere is, of course, expected to be somewhat non-spherical and the radiation field anisotropic). In line with this assumption, we approximate the photospheric temperature as 
\begin{equation}
    T_{\rm ph} \approx f_{\rm col} \left(L_{\rm Edd}/4\pi \sigma R_{\rm ph}^{2}\right)^{1/4}
\end{equation}
We fix the colour temperature correction factor, $f_{\rm col}$ = 2, for simplicity (strictly speaking, this value will have its own dependence on $T_{\rm ph}$ and therefore on accretion rate). At this point, the radiation field impinging on the outer disc is completely specified.

\subsection{The formation of irradiation-driven thermal disc winds}

The existence of an intense source of radiation at the centre of an extended accretion disc raises the possibility of driving a second wind, at large radii. The driving mechanism for such a wind is different to that which produces the central quasi-spherical outflow in ULXs (which is instead driven by radiation pressure); in this case, mass loss results from heating of the surface layers of the accretion disc. If the gas reaches a temperature such that the thermal velocity of the gas particles exceeds the local escape velocity then the gas is able to expand away from the disc and form a wind (\citealt{Begelman83}). 

Two points are key in assessing whether irradiation will lead to mass loss at all and, if so, how strong it will be. Firstly, the maximum temperature in the irradiated disc atmosphere determines whether and where in the disc, thermal driving can occur. Secondly, the strength of mass loss depends on whether material in the atmosphere reaches this temperature gradually or explosively, i.e. whether the irradiation produces a thermal instability in the atmosphere. Both of these points can be explored by considering the ``stability curve" for the irradiated material.

\begin{figure}
    \includegraphics[width=\columnwidth]{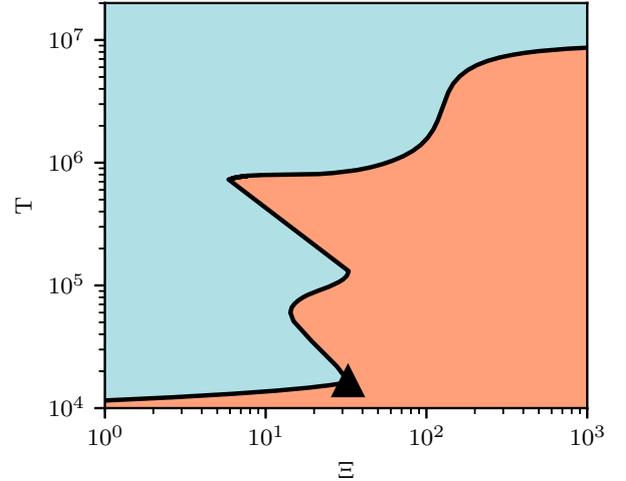}
    \caption{A stability curve for gas illuminated by a 
    source radiating as a black body of temperature $1\times10^7$K. Gas in the red portion of the figure is heating, in the blue portion cooling, and gas on the black line is stable. The symbol shows the location of $\Xi_{\rm cool,max}$. }
    \label{fig:stab1}
\end{figure}

Figure \ref{fig:stab1} shows such a stability curve for gas illuminated by a $10^7$K black body. More specifically, it shows the equilibrium temperature, $T_{\rm eq}$, at which heating balances cooling, as a function of pressure ionization parameter $\Xi$, given by
\begin{equation}
\Xi=\frac{P_{\rm rad}}{P_{\rm gas}}=\frac{L}{4\pi R^2cP_{\rm gas}}.
\label{Equation:press_ion_param}
\end{equation}
Here, $L$ is the luminosity of the central source, in this case the photosphere (and we therefore assume $L = L_{\rm Edd}$), $P_{\rm gas}$ is the gas pressure, $P_{\rm rad}$ is the radiation pressure and 
$R$ is the distance between the source of radiation and an element of the disc surface. 

All regions of the $T_{\rm eq} (\Xi)$ curve with positive slope correspond to thermally stable configurations. Here, any increase in the heating rate is balanced by an increase in cooling rate (and vice versa). These stable regions occur at low temperatures (representing the disc atmosphere) and at high temperatures (where the temperature is eventually set by Compton heating and cooling). Between these regimes, there is an unstable region comprising a series of ``switchbacks". Negative slopes in such figures indicate the onset of thermal instability (such that an increase in the heating rate is not immediately balanced by an increase in the cooling rate). Whilst thermal instability is not a pre-requisite for thermal winds, the rapid heating it causes is important for driving the fast and strong winds that can have a sizeable impact on the accretion rate feeding the inner regions (see \citealt{Higginbottom_Proga2015}). Following \cite{Higginbottom17}, we indicate the position of $\Xi_{\rm cool, max}$, the point at which gas, heating up along the `cool' branch, becomes thermally unstable and heats up very rapidly towards the `hot' branch. 

\begin{figure}
    \includegraphics[width=\columnwidth]{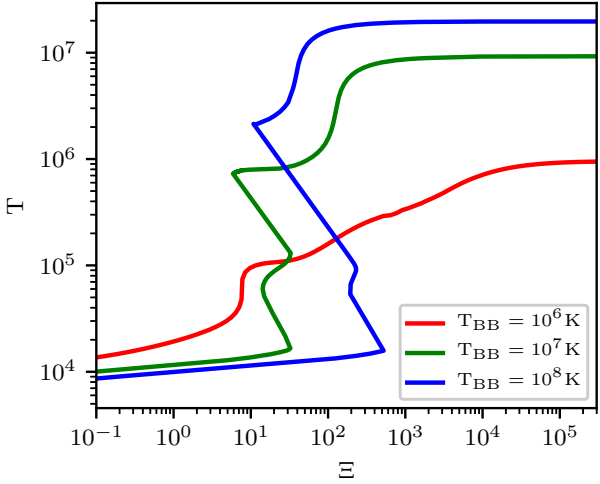}
    \caption{Set of stability curves for gas illuminated by blackbodies of temperature $1\times10^6$K, $1\times10^7$K and $1\times10^8$K. The $1\times10^6$K curve does not exhibit unstable heating, whilst for the $1\times10^7$K curve, thermal instability sets in at $\Xi_{\rm cool,max}\sim40$ and for $1\times10^8$K the onset is at $\Xi_{\rm cool,max}\sim500$.}
    \label{fig:stab2}
\end{figure}

The stability curve -- and whether a region of instability bridges the two stable branches -- is a sensitive function of the irradiating SED. In Figure \ref{fig:stab2} we plot a family of stability curves for various black body temperatures, which indicate that the transition from globally stable to partially unstable  occurs around $T_{\rm BB} \simeq 10^{6}$ K. In the case of ULXs, the irradiating SED is a function of the accretion rate, compact object mass and, in the case of neutron stars, the dipole magnetic field. As we shall see, physically relevant combinations of these parameters can give rise to $T_{\rm BB} > 10^{6}$~K. Such systems are likely to suffer strong, irradiation-driven mass loss, provided that their discs are large enough. 


Sufficiently high up in the irradiated disc atmosphere, heating and cooling will generally be dominated by Compton processes. The maximum temperature reached in the outer disc atmosphere -- at any given radius -- is therefore the inverse Compton temperature, $T_{\rm IC}$.
This is {\textrm only} a function of the {\textrm shape} of the irradiating SED (c.f. Figure \ref{fig:stab2}, where each of the stability curves asymptotes to a different value of $T_{\rm IC}$ as $\Xi \rightarrow \infty$.) Thus we can define a unique radius for any given system beyond which the characteristic thermal speed of particles at $T_{\rm IC}$ will exceed the escape speed. This is the so-called (inverse) Compton radius, $R_{\rm IC}$:
\begin{equation}
R_{\rm IC}=\frac{GM\mu m_{\rm H}}{k_{\rm B}T_{\rm IC}} =9.8 \times 10^{9} \left(\frac{T_{\rm IC}}{10^{8} K}\right)^{-1}\left(\frac{M}{M_{\odot}}\right) ~~~ {\rm cm}
\end{equation}
where $k_{\rm B}$ is Boltzmann's constant, $\mu$ is the mean gas mass in units of hydrogen mass ($m_{\rm H}$), and $M$ is the mass of the compact object. In fact, significant mass loss is already expected for $R \gtrsim 0.1R_{\rm IC}$ \citep{1986ApJ...306...90S}. 

We proceed to parameterise the rate of mass loss in the thermal wind as $\dot{m}_{\rm th}$, the value of which requires in-depth RHD simulations (which we reserve for a follow-up work). We define $\dot{m}_{\rm 0}$ to be the mass transfer rate from the secondary star in units of the Eddington accretion rate, such that the resulting accretion rate at $r_{\rm sph}$ is then $\dot{m} = \dot{m}_{\rm 0} - \dot{m}_{\rm th}$.
In the case where thermal winds are launched, such that $\dot{m} < \dot{m}_{\rm 0}$, $r_{\rm sph}$ and $r_{\rm ph}$ will move inwards relative to the case where $\dot{m}_{\rm th} = 0$. Should $\dot{m}_{\rm th}$ itself be modulated over time, we can make broad predictions for what should happen to those observables with a strong dependence on $\dot{m}$. Notably, as $\dot{m}$ decreases due to mass loss at larger radii, the photosphere will retreat to smaller radii leading to an increase in $T_{\rm ph}$; in turn this will change the position of $R_{\rm IC}$ and the rate of mass loss $\dot{m}_{\rm th}$. It is therefore quite plausible that mass loss via a thermal wind could lead to long timescale changes in the system or limit cycles, depending on $\dot{m}_{\rm th}$ and $\dot{m}_{\rm 0}$.


\subsection{Mass loss rate in the thermal wind} 

To explore the scenarios resulting from thermal wind mass loss in ULXs, it is clearly important to obtain the dependence of $\dot{m}_{\rm th}$ on various system parameters. We create our framework using the preceding formulae for the location of the respective radii ($r_{\rm sph}$, $r_{\rm ph}$, $r_{\rm IC}$) and the temperature of the photosphere $T_{\rm ph}$. 
$T_{\rm IC}$ is related to the temperature at the photosphere of the wind by $T_{\rm IC} \approx 0.675T_{\rm ph}$ (\citealt{1996ApJ...461..767W}) such that

\begin{equation}
T_{\rm IC} \approx 6.2 \times 10^{6} \left(\frac{\xi\beta}{\epsilon}\right)^{1/2}\left(\frac{M}{M_{\odot}}\right)^{-1/4}\dot{m}^{-3/4} ~~~ {\rm K}
\end{equation}

\noindent which then implies

\begin{equation}
    R_{\rm IC} \approx 1.6 \times 10^{11} \left(\frac{\xi\beta}{\epsilon}\right)^{-1/2}\left(\frac{M}{M_{\odot}}\right)^{5/4}\dot{m}^{3/4} ~~~ {\rm cm}
\end{equation}


The model of \cite{1996ApJ...461..767W}, based on \cite{1983ApJ...271...70B}, assumes the irradiation of the outer disc is from a point source. Assuming a roughly spherical photosphere, it is reasonable to assume that the wind will start to resemble a point source for $r_{\rm IC}/r_{\rm ph} >$ 100 (where $r_{\rm IC}$ is in units of $r_{\rm in}$), which occurs for $\dot{m} <$ 5,000 (for the more stringent case of a neutron star). Hereafter we assume that the point source approximation is reasonable, but note that locating the photosphere requires a full RMHD treatment.

Following \cite{1996ApJ...461..767W}, we write the mass-loss rate in the thermal wind as
\begin{equation}
    \dot{M}_{\rm th} = \int_{R_{\rm ph}}^{R_{\rm disc}} \mathcal{\dot{M}}  \, \times \,2 \times \left(2\,\pi\,R\,dR\right),
\end{equation}
where $\mathcal{\dot{M}}$ is the mass-loss rate per unit area from the disc. A characteristic value for this is given by $\mathcal{\dot{M}}_{\rm ch} = p_{\rm 0} / c_{\rm ch}$, where $p_{\rm 0}$ and $c_{\rm ch}$ are the pressure and sound speed respectively, at the top of the thermally stable part of the atmosphere. 
$\mathcal{\dot{M}}$ can then be written in terms of $\mathcal{\dot{M}}_{\rm ch}$ as
\begin{align}
    \mathcal{\dot{M}} = \mathcal{\dot{M}}_{ch}
\left\{\frac{1 + \left[\left(0.125L/L_{\rm cr} + 0.00382\right)/X\right]^2}{1 + \left[\left(L/L_{\rm cr}\right)^4\left(1+262X^2\right)\right]^{-2}}
\right\}^{1/6}
\times \notag
\\ 
\exp{
\left\{
\frac{-\left[1-\left(1+0.25X^{-2}\right)^{-1/2}\right]^2}{2X}
\right\}
},
\end{align}
where $X = R/R_{\rm IC}$, and $L_{\rm cr}$ is a critical luminosity defined as
\begin{equation}
    L_{\rm cr} = 2.88 \times 10^{-2} \left(\frac{T_{\rm IC}}{10^{8}{\rm K}}\right)^{-1/2} L_{\rm Edd}.
\end{equation}

In order to actually evaluate Equation~7, we still need expressions for $p_{\rm 0}$ and $c_{\rm ch}$. Following \cite{1996ApJ...461..767W}, we write  $p_{\rm 0}$ as
\begin{align}
p_0 = 1.1 \times 10^{5} \,\,
\left(\frac{L}{L_{\rm Edd}}\right)
\left(\frac{M}{\mathrm{M_{\odot}}}\right)^{-1}
\left(\frac{T_{\rm IC}}{10^8\,\mathrm{K}}\right)^{2}
\left(\frac{\Xi_{\rm c,max}}{40}\right)^{-1} \notag \\ 
\times \left(\frac{R}{R_{\rm IC}}\right)^{-2}
\, \mathrm{ergs~cm^{-3}},
\end{align}
where $\Xi_{\rm cool,max}$ is the pressure ionisation parameter at the top of the thermally stable part of the atmosphere. To calculate the (isothermal) sound speed at this location, we use the standard approximation (\citealt{FKR}), ${\rm c_{\rm ch}} \simeq 10~\mathrm{km~s^{-1}} \sqrt{T_{\rm ch}/10^4\,\mathrm{K}}$, where $T_{\rm ch}$ is the local gas temperature. This, in turn, can be estimated as \citep{1996ApJ...461..767W}
\begin{equation}
T_{\rm ch} = T_{\rm IC}
\left(\frac{L}{L_{\rm cr}}\right)^{2/3}
X^{-2/3}.
\end{equation}
Putting all this together yields
\begin{equation}
{\rm c_{ch}} \simeq 10^{8} \left(\frac{T_{\rm IC}}{10^8\,\mathrm{K}}\right)^{1/2}
\left(\frac{L}{L_{\rm cr}}\right)^{1/3}
X^{-1/3}
\mathrm{cm~s^{-1}}.
\end{equation}



\noindent  Combining our expressions for $p_0$, $L_{\rm cr}$ and $c_{\rm ch}$, we find
\begin{align}
\mathcal{\dot{M}}_{ch} = 3.4 \times 10^{-4} \left(\frac{L}{L_{\rm Edd}}\right)^{2/3} \left(\frac{M}{\mathrm{M_{\odot}}}\right)^{-1}\left(\frac{T_{\rm IC}}{10^8\,\mathrm{K}}\right)^{4/3} \notag \\ 
\times \left(\frac{\Xi_{\rm c,max}}{40}\right)^{-1} X^{-5/3} \, \mathrm{g~s^{-1}~cm^{-2}}.
\end{align}

\noindent Setting $X_{\rm ph} = \frac{R_{\rm ph}}{R_{\rm IC}}$ and $X_{\rm disc} = \frac{R_{\rm disc}}{R_{\rm IC}}$ then yields the mass-loss rate in the thermal wind:

\begin{align}
\dot{M}_{\rm th} 
& \simeq \int_{R_{\rm ph}}^{R_{\rm disc}} \mathcal{\dot{M}}  \, \times \,2 \times \left(2\,\pi\,R\,dR\right) \notag \\ \notag \\
& = 3.4 \times 10^{-4}\,\,(4\pi)\,\,R_{\rm IC}^2
\left(\frac{L}{L_{\rm Edd}}\right)^{2/3}
\left(\frac{M}{\mathrm{M_{\odot}}}\right)^{-1}
\left(\frac{T_{\rm IC}}{10^8\,\mathrm{K}}\right)^{4/3} \notag \\
& \left(\frac{\Xi_{\rm c,max}}{40}\right)^{-1}
 \times \int_{X_{\rm ph}}^{X_{\rm disc}}
\left\{\frac{1 + \left[\left(0.125L/L_{\rm cr} + 0.00382\right)/X\right]^2}{1 + \left[\left(L/L_{\rm cr}\right)^4\left(1+262X^2\right)\right]^{-2}}
\right\}^{1/6} \notag \\ 
&\times \exp{
\left\{
\frac{-\left[1-\left(1+0.25X^{-2}\right)^{-1/2}\right]^2}{2X}
\right\}
}\times X^{-2/3} dX ~\mathrm{g~s^{-1}}.
\end{align}

\noindent Finally, after simplifying and substituting for $R_{\rm IC}$ using equation 3, we obtain

\begin{align}
\dot{M}_{\rm th} 
& = 4.1 \times 10^{17}
\left(\frac{L}{L_{\rm Edd}}\right)^{2/3}
\left(\frac{M}{\mathrm{M_{\odot}}}\right)
\left(\frac{T_{\rm IC}}{10^8\,\mathrm{K}}\right)^{-2/3}
 \left(\frac{\Xi_{\rm c,max}}{40}\right)^{-1} \notag \\
& \times \int_{X_{\rm ph}}^{X_{\rm disc}}
\left\{\frac{1 + \left[\left(0.125L/L_{\rm cr} + 0.00382\right)/X\right]^2}{1 + \left[\left(L/L_{\rm cr}\right)^4\left(1+262X^2\right)\right]^{-2}}
\right\}^{1/6} \notag \\ 
&\times \exp{
\left\{
\frac{-\left[1-\left(1+0.25X^{-2}\right)^{-1/2}\right]^2}{2X}
\right\}
}\times X^{-2/3} dX ~\mathrm{g~s^{-1}}.
\label{equation:massloss}
\end{align}

\noindent which needs to be calculated numerically. As discussed earlier, $\Xi_{\rm cool,max}$ depends on the shape of the irradiating spectrum, however, since we are assuming this to be a simple blackbody and are ignoring optical depth effects, this is just a function of $T_{\rm IC}$.

\begin{figure}
    \includegraphics[width=\columnwidth]{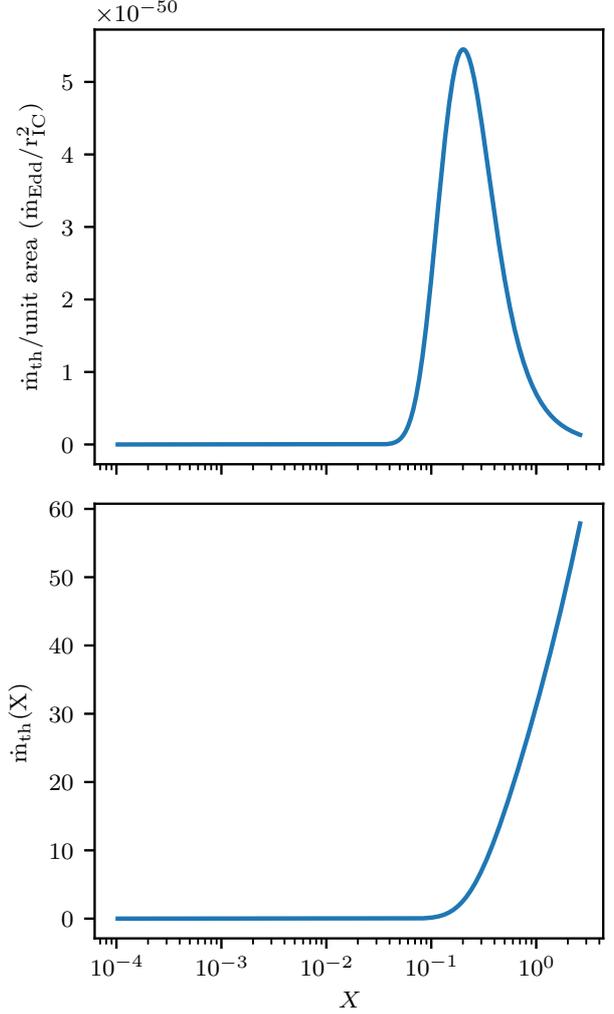}
    \caption{An example of the mass-loss rate per unit area and cumulative mass-loss rate for a thermally driven wind in a system with a photospheric temperature of $3.6\times10^6$K ($\dot{m}_{0} = 10$, M = 10 M$_{\odot}$).}
    \label{fig:mdot_vs_r}
\end{figure}

The upper panel of Figure \ref{fig:mdot_vs_r} shows the form of the integrand in equation \ref{equation:massloss} as a function of $X$ for example values of $\dot{m}_{0} = 10$ and M = 10 M$_{\odot}$. We can clearly see how there is essentially zero mass loss for radii interior to about $0.1R_{\rm IC}$. The lower panel shows the integrated mass-loss due to the thermal wind out to a given value of $X$ and demonstrates how large the integrated mass loss can be.

\section{Results}

\subsection{The impact of disc winds on ULXs}

\begin{figure}
    \includegraphics[width=\columnwidth]{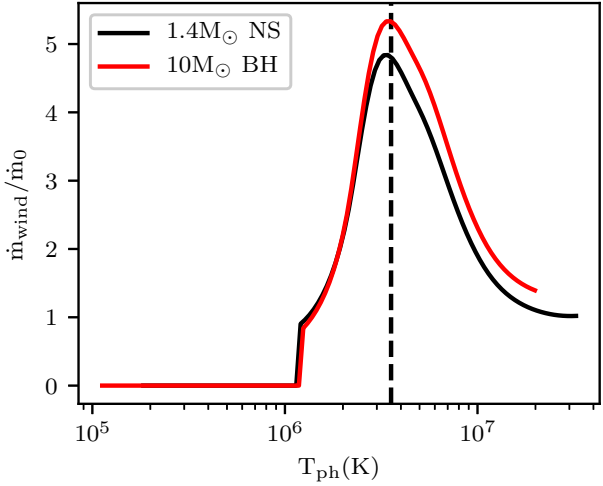}
    \caption{The wind efficiency as a function of photospheric temperature for a $1.4$ M$_{\odot}$ neutron star and a 10 M$_{\odot}$ black hole. The dashed line shows the temperature
    at which Figure \ref{fig:mdot_vs_r} is made.}
    \label{fig:eff}
\end{figure}

\begin{figure}
    \includegraphics[width=\columnwidth]{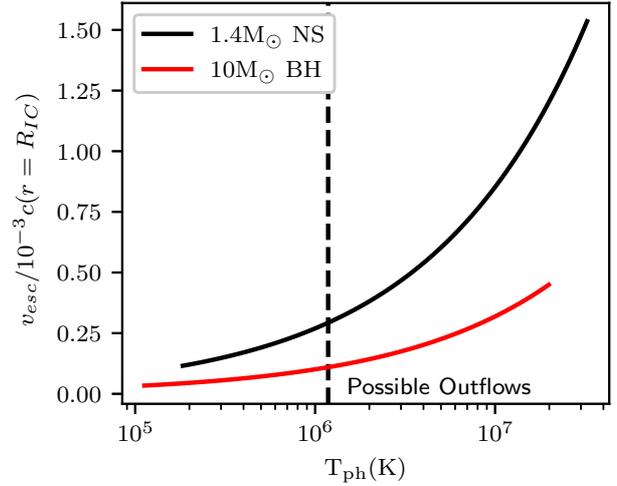}
    \caption{The escape velocity (in units of $10^{-3}$c) at the inverse Compton radius for a range of photospheric temperatures. The vertical dashed line shows the minimum photospheric temperature required to launch a wind.}
    \label{fig:vkep}
\end{figure}

The model described in the previous section predicts the mass loss rate via a thermal wind, based upon the temperature of the photosphere which, in the absence of strong dipole magnetic fields depends only on the accretion rate into the disc and the mass of the compact object. Figure \ref{fig:eff} shows the wind efficiency (i.e. the ratio of mass-loss rate via a thermal wind compared to the initial accretion rate $\dot{m}_0$). An efficiency of greater than 1 indicates it is highly likely that the accretion process would be heavily disrupted by the presence of a thermal wind, leading to a modulation in the observed system properties.
Figure \ref{fig:eff} shows that, wherever $T_{\rm ph}$ is high enough for a thermal instability to exist, the wind efficiency is indeed greater than 1. We also show the escape velocity at $R_{\rm IC}$ for the case of a black hole and neutron star in Figure \ref{fig:vkep}.

The presence of thermal winds with a mass loss rate defined by equation \ref{equation:massloss}, will affect the location of $r_{\rm sph}$. The change in location of $r_{\rm sph}$ will occur after the time it takes the change in accretion rate to propagate down to this radius, which is approximately the viscous timescale around $R_{\rm IC}$. Given the complexity regarding the mass loss as a function of radius, we will limit ourselves to exploring only the regions of system parameter space which we predict to contain unstable ULXs, and in future will explore how these systems appear as a function of time.

\begin{figure}
    \includegraphics[width=\columnwidth]{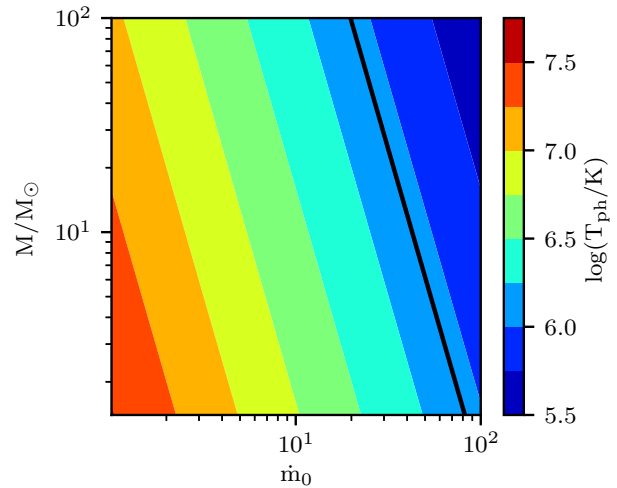}
    \caption{The photospheric temperature as a function of mass and 
    accretion rate (in the absence of strong dipole magnetic fields in the case of neutron stars). The black line divides systems with a photospheric temperature which 
    may lead to a significant disc wind (to the left of the line) from those which are too cool to 
    induce a thermal instability (to the right of the line)} 
    \label{fig:temperature}
\end{figure}

Figure \ref{fig:temperature} shows the values of $T_{\rm ph}$ for a range of compact object masses and $\dot{m}_{0}$. The black line shows $T_{\rm ph}=1.2\times10^6$, the point at which thermal instability, and therefore powerful thermal winds, become possible. Therefore, systems to the left hand side of this figure (lower mass and/or lower accretion rates) are the ones for which thermal winds are likely to have a significant impact on the system's behaviour.

\subsection{The impact of magnetic fields}

The above results are approximately appropriate for discs which extend from the ISCO to $r_{\rm out}$. However, the discovery of pulsating ULXs indicates the presence of dipole magnetic fields which, if strong (typically $\gtrsim$ 10$^{9}$ G), can truncate the disc at the magnetospheric radius (\citealt{Davidson_Ostriker_Rm}), with the field strength typically inferred from assuming the source to be close to spin equilibrium (e.g. \citealt{NSULX_Bachetti_2014}) or, more rarely, from CRSFs (\citealt{2018_Brightman_CRSF,Middleton_Brightman_2019_M51}). Whilst there is still no entirely unambiguous estimate, values for the dipole field appear to be around those of Galactic HMXBs (10$^{12}$ - 10$^{13}$ G (\citealt{King_2017_Pulsating_ULXs}), potentially with strong multipolar components (e.g. \citealt{Israel_2017_NSULX_5907, Middleton_Brightman_2019_M51}). 

The presence of a dipole field has no effect on the position of the spherisation radius (regardless of whether this is actually found in the disc) but can affect the location of the photospheric radius as a consequence of the optical depth through the wind being an integrated quantity (\citealt{Poutanen_2007_ln}). Following \cite{2019_Vasilopoulos_MNRAS.488.5225V} (see also \citealt{Middleton_2019_Accretion_plane}), we can determine the position of $r_{\rm ph}$ from

\begin{equation}
    r_{\rm ph} \approx \frac{\tau_{0}}{\beta}\frac{\dot{m}_{0}}{\sqrt{r_{\rm sph}}}\left(r_{\rm sph} - r_{\rm M}\right)
\end{equation}

\noindent, however, to be consistent with the formula assuming no truncation (equivalent to where $r_{\rm sph}\gg r_{\rm M}$), we re-write this as

\begin{equation}
    r_{\rm ph} \approx \frac{3\epsilon_{\rm w}}{\beta\xi}\frac{\dot{m}_{0}}{\sqrt{r_{\rm sph}}}\left(r_{\rm sph} - r_{\rm M}\right)
\end{equation}





\noindent in units of $r_{\rm in}$, and where $r_{\rm M}$ is the magnetospheric radius in a super-critical disc, which we assume to be

\begin{equation}
    R_{\rm M} \approx 2.9\times10^{8}\dot{M}_{17}^{-2/7}m_{\rm NS}^{-1/7}\mu_{30}^{4/7}~~~~~{\rm cm}
\end{equation}



\noindent where $\dot{M}_{17}$ is the mass accretion rate in units of 10$^{17}$ g/s, $m_{\rm NS}$ is the neutron star mass in solar units and $\mu_{30}$ is the magnetic dipole moment in units of 10$^{30}$ Gcm$^{3}$. In a classical super-critical flow (i.e. one where $r_{\rm sph} > r_{\rm M}$), the mass accretion rate scales linearly with radius (e.g. \citealt{Shakura_1973,Poutanen_2007_ln}) such that $\dot{M}_{17} = \dot{M}_{0,17}R/R_{\rm sph}\approx R/R_{\rm in}$; this then allows us to estimate the position of $R_{\rm M}$, independent of accretion rate (see \citealt{King_2017_Pulsating_ULXs,Middleton_2019_Accretion_plane}).

\begin{figure}
    \includegraphics[width=\columnwidth]{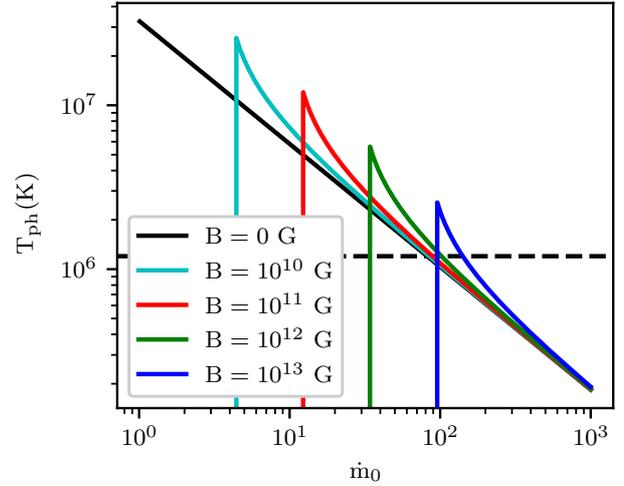}
    \caption{The photospheric temperature of a 1.4 M$_{\odot}$ neutron star ULX as a function of $\dot{m}_0$ for a range of dipole magnetic field strengths. As the magnetic field acts to reduce the radius of the photosphere (see equation 16), the photospheric temperature is higher compared to the case with no (or a weak) dipole field, thereby pushing the outer disc closer to thermal instability, indicated by being above the horizontal dashed line. The vertical drop indicates where $r_{\rm sph} = r_{\rm M}$.} 
    \label{fig:mag_field}
\end{figure}

Figure \ref{fig:mag_field} shows the resulting effect of dipole magnetic field strength on the photospheric temperature, with the horizontal dashed line showing the temperature above which a strong thermal wind might exist. The vertical drop in the figure indicates the point at which $r_{\rm M} = r_{\rm sph}$. 
Although the presence of a dipole field can clearly make neutron star ULX systems more thermally unstable (indicated by the difference between the solid black line and the coloured lines), this is only relevant for high field strengths ($>$ 10$^{12}$ G) and high accretion rates ($\dot{m} >$ 100). For completeness we also explore the form of $r_{\rm M}$ provided by \cite{2019Chashkina}, and obtain consistent results.

\subsection{The population of thermally unstable ULXs}


In order to explore the types of ULX which may experience significant thermal mass loss, we utilise the population produced by the binary population synthesis code {\sc startrack} as presented in \cite{Wiktorowicz_2017_Origin}. This code evolves binaries across a range of model parameters assuming a Kroupa IMF (\citealt{Kroupa03}), and metallicities between Z = 0.001 and 0.1. We assume a constant star formation history (i.e. we draw systems from any stage in their evolution) as we are only interested in the variety of ULXs which could be produced, rather than the content of any particular galactic environment.


The ULXs formed by the code include neutron stars and black holes, experiencing a mixture of nuclear and thermal timescale Roche lobe overflow. Of key importance is whether 0.1$R_{\rm IC}$ lies within the outer radius of the disc in these systems (otherwise we assume that strong thermal winds are not driven). We assume the outer edge of the disc, R$_{\rm out} <~$R$_{\rm tidal} \approx 0.9 $R$_{1}$ (\citealt{FKR}), where R$_{1}$ is the size of the primary's Roche lobe given by the standard formula from \cite{Eggleton83}: 

\begin{equation}
    R_{1} = \frac {0.49aq^{2/3}}{0.6q^{2/3} + {\rm ln}(1+q^{1/3})}
\end{equation}

\noindent where, $a$ is the binary separation and, as opposed to the usual case, q = $M_{1}/M_{2}$ (with values obtained for each system directly from the simulation). 
Using the compact object mass, accretion rate (and assuming $r_{\rm in}$ = 6 R$_{\rm g}$), we obtain the irradiating SED for each ULX in the sample with a luminosity of 1.3$\times 10^{38}M_{1}$ erg/s (with $M_1$ in units of M$_{\odot}$). We ignore the presence of magnetic fields as we have seen that the field makes very little difference to the triggering of strong thermal winds (and the irradiating luminosity is assumed independent of magnetic field strength). Figure \ref{fig:population} shows all of the systems produced in the simulation for which 0.1R$_{\rm IC} < ~$R$_{\rm out}$, and which could therefore potentially drive thermal winds. Our choice of constants ($\beta, \xi$ and $\epsilon_{\rm w}$), implies that
$r_{\rm ph} < r_{\rm sph}$ for $\dot{m}_{0} < 1$, i.e. for systems which are not super-critical and are therefore not shown in this plot. In Figure \ref{fig:period} we also plot contours enclosing the orbital period vs mass accretion rate space for those thermally unstable neutron star and black hole systems.

\section{Discussion \& Conclusion}

We have explored the conditions required for driving powerful thermal winds (following \citealt{Begelman83}) from the outer, non-super-critical discs of ULXs when irradiated by an assumed isotropically emitting photosphere of a radiatively driven wind (e.g. \citealt{Yao_Feng2019_discirrad}). As indicated in Figure \ref{fig:temperature}, there is a substantial area of parameter space (typically $\dot{m} < 100$) where thermal winds should have a major impact on the flow of material down to $r_{\rm sph}$ (i.e. the mass loss rate is $\gtrsim$ the mass feeding rate from the secondary). Although dipole fields can affect the location of the photosphere (by truncating the disc at $r_{\rm M}$ and thereby changing the integrated optical depth through the wind), they tend not to push previously stable systems into instability (unless at high accretion rate and with a high dipole field strength).

\begin{figure}
    \includegraphics[width=\columnwidth]{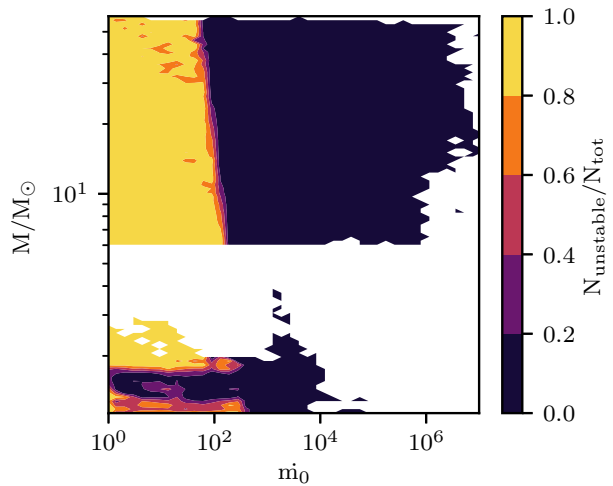}
    \caption{The proportion of systems (parameterised by N$_{\rm unstable}$/N$_{\rm tot}$) extracted from a \textsc{startrack} 
    simulation, where unstable behaviour is predicted based upon 
    photospheric temperature and outer disc size. White areas are where 
    there are no systems present in the simulation.}
    \label{fig:population}
\end{figure}
\begin{figure}
    \includegraphics[width=\columnwidth]{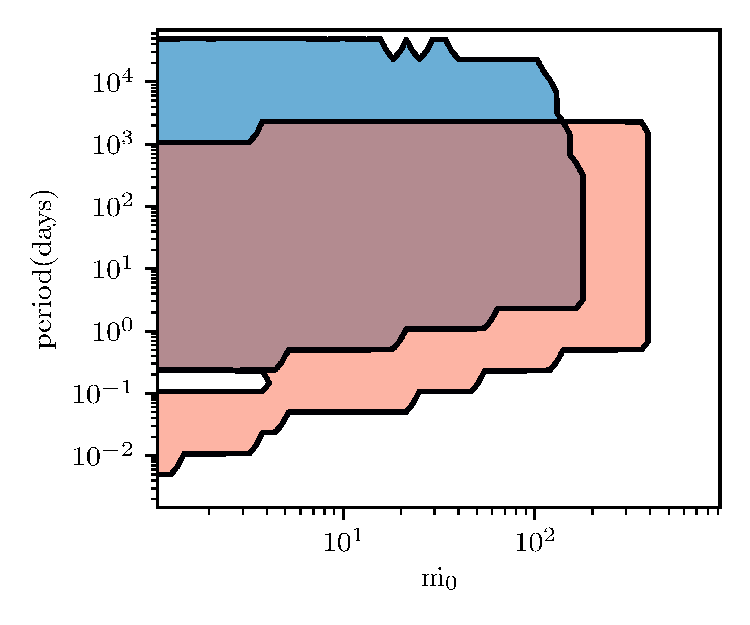}
    \caption{Enclosed regions for the orbital period and mass transfer rate of those unstable systems shown in Figure \ref{fig:population}; black hole systems are found in
    the blue area, neutron stars in the foreground 
    red area.}
    \label{fig:period}
\end{figure}

We explore the binary parameter space (via simulations using {\sc startrack}: \citealt{Wiktorowicz_2017_Origin}) to locate systems where powerful thermal winds are predicted, based on the criteria that the outer disc is large enough to support their presence (i.e $R_{\rm out} > 0.1 R_{\rm IC}$) and where $r_{\rm ph} > r_{\rm sph}$ (to ensure we are considering emission only from the wind photosphere). We find that both thermally unstable black hole and neutron star ULX systems tend to have accretion rates $\dot{m}_0 <$ 100 (Figure \ref{fig:population}). In Figure \ref{fig:period}, we plot the orbital periods and accretion rates for those unstable systems shown in Figure \ref{fig:population}. Estimates for $\dot{m}_0$ in neutron star ULXs (e.g. \citealt{King_Lasota_2020_PULX_iceberg_emerges}) and black hole candidate ULXs (e.g. \citealt{Middleton_2019_Accretion_plane}), and orbital periods of those known neutron star ULXs ($\lesssim$ 10s of days e.g. \citealt{NSULX_Bachetti_2014,Israel_2017_NSULX_5907,Fuerst_P13_2021}) indicate that some well-studied ULXs should lie within the unstable regions of Figures \ref{fig:population} and \ref{fig:period}. 
Under the assumption that the photosphere emits isotropically (rather than being directed away from the outer disc by the wind), such thermally unstable systems should undergo substantial changes in their observed properties over time. 

It is rather noticeable that the bright ULXs tend not to switch off for extended periods (although faint states are observed in some ULX pulsars, e.g. \citealt{Fuerst_P13_2021}). However, it has also been suggested that optical emission from such systems is driven, at least in part by irradiation of the outer disc (e.g. \citealt{Sutton2014_irraddisc}). Given the ease with which thermal winds should be driven in such systems, this would seem to imply that the irradiation is sub-Eddington, such that the mass loss rate (via equation 13, see also \citealt{2019MNRAS.484.4635H}) is lower, and the accretion flow is not terminated but instead only somewhat diminished. Such sub-Eddington irradiation may be a natural consequence of anisotropy of an Eddington-emitting photosphere, itself due to the inhomogeneity of winds, which instead have a complex density/velocity structure (e.g. \citealt{Takeuchi2013}). It is possible to reconcile lower amounts of irradiation with observation if the optical emission originates from the photosphere of the wind (should the accretion rate be large, \citealt{Poutanen_2007_ln}), from irradiation of the secondary star (e.g. \citealt{Motch2014}), or together in some combination with the irradiated outer disc (e.g. \citealt{Copperwheat2007,Patruno2008}). Exploring this requires the photosphere of the wind to be extracted from simulation, and radiative transfer performed accounting for the outflowing nature of mass through the photosphere. In addition to this, it may be important to establish what fraction of harder X-rays -- created in in the innermost regions -- are able to scatter in optically thin components of the wind and contribute to the irradiation of the outer disc (the temperature being high enough to readily induce thermal winds). Whilst a detailed study of this type is beyond the scope of this initial paper, it will be the focus of future work. 

In the case where the irradiation is indeed sub-Eddington and non-terminating mass loss induced, we would expect other observable quantities such as the luminosity, spectrum (e.g. the temperature at the spherisation radius), and potentially precession timescale (if sensitive to the accretion rate: \citealt{Middleton_2019_Accretion_plane}) to change and limit cycles potentially induced. Such a process may already help explain the long timescale changes seen in systems such as NGC 1313 X-1 (the change in flaring timescales - see, e.g. \citealt{Walton_1313_2020}). Conversely it is intriguing that the highly super-critical Galactic system, SS433 (see \citealt{Fabrika_2004_SS433} for a review), is relatively stable on long timescales, implying its disc is either too small to support thermal wind production or the Eddington-scaled accretion rate is too high to readily drive such thermal winds (indeed, the wind mass-loss rate in SS433 implies $\dot{m}_{0} >$~100 for a stellar mass black hole: \citealt{Shklovsky_ss433,Fuchs2006}). Should we find that thermal wind-induced changes in accretion rate do appear to match observation, the implication is that the outer disc may indeed be efficiently irradiated (consistent with some portion of the optical emission then being produced in the outer disc), which might in turn lead to the generation of radiative warps (as seen in Her X-1: \citealt{Petterson_1977_Her_X1_Precession}) which may provide a mechanism for producing some of the observed super-orbital periods (see \citealt{Middleton_2018_Lense_Thirring} for a discussion of the various mechanisms as applied to ULXs).

In Figure \ref{fig:vkep} we show the escape velocity at R$_{\rm IC}$ (for both a black hole and a neutron star). Assuming this value to be a reasonable indicator of the actual velocity, the thermal winds we predict, will be ejected at speeds orders of magnitude less than those of the fast winds revealed by high resolution X-ray spectroscopy (e.g. \citealt{Pinto_1313_2020}). However, such relatively slow-moving thermal winds may provide a natural explanation for observations of rest-frame emission lines (e.g. \citealt{Pinto_2016Natur.533...64P}) as such winds are unlikely to obscure the inner accretion flow but will emit as the gas cools and recombines. We will explore this explicitly via simulations in a follow-up work.


The next step to obtaining a clearer understanding of the impact of thermal winds on ULXs will be to perform extensive RHD simulations to better understand the mass loss rate as well as to explore the emergence of radiation at the photosphere (and thereby evaluate the irradiation of the outer disc). Going forwards, we will also explore the time dependence of the mass loss and its impact on the disc in detail.



\section*{Acknowledgements}

We thank the anonymous referee for their useful comments and suggestions, and Chris Done for valuable discussion.

\section*{Data Availability}

Data from the running of the binary population synthesis code can be found at https://universeathome.pl/universe/bhdb.php. 




\bibliographystyle{mnras.bst}
\bibliography{bibliography.bib}  





\bsp	
\label{lastpage}
\end{document}